\documentclass[3p,times,twocolumn]{elsarticle}

\usepackage{ecrc}
\usepackage{amsmath, setspace,verbatim}
\usepackage{color}
\usepackage{xcolor}
\usepackage{wrapfig}
\usepackage{etoolbox}
\usepackage{relsize}
\usepackage{multirow}

\usepackage{graphicx} 
\usepackage{array}

\usepackage{hyperref} 
\usepackage{float}
\usepackage{soul}

\volume{00}

\firstpage{1}

\journalname{Nuclear and Particle Physics Proceedings}

\runauth{T. Hurth et al.}

\jid{nppp}

\jnltitlelogo{Nuclear and Particle Physics Proceedings}

\usepackage{amssymb}
\biboptions{sort&compress}

\usepackage[figuresright]{rotating}

\begin{document}

\begin{frontmatter}

\dochead{}

\title{New global fits to $b \to s$ data with all relevant parameters}


\author{Tobias Hurth\tnoteref{label1}}
\tnotetext[label1]{Speaker \\ \hfill{\tt CERN-TH-2018-244, IPM/P.A-514, MITP/18-120}}
\address{PRISMA Cluster of Excellence \& Institute for Physics (THEP)\\
Johannes Gutenberg University, D-55099 Mainz, Germany}
%\ead{tobias.hurth@cern.ch}
\author{Alexandre Arbey}
\address{Univ
 Lyon, Univ Lyon 1, ENS de Lyon, CNRS, Centre de Recherche Astrophysique de Lyon\\
UMR5574, F-69230 Saint-Genis-Laval, France;\\
Theoretical Physics Department, CERN, CH-1211 Geneva 23, Switzerland}
%\ead{alexandre.arbey@ens-lyon.fr}
\author{Farvah Mahmoudi}
\address{Univ
 Lyon, Univ Lyon 1, ENS de Lyon, CNRS, Centre de Recherche Astrophysique de Lyon\\
UMR5574, F-69230 Saint-Genis-Laval, France;\\
Theoretical Physics Department, CERN, CH-1211 Geneva 23, Switzerland}
%\ead{nazila@cern.ch}
\author{Siavash Neshatpour}
\address{School
 of Particles and Accelerators, Institute for Research in Fundamental Sciences (IPM)\\
P.O. Box 19395-5531, Tehran, Iran}
%\ead{neshatpour@ipm.ir}

\vspace{-0.8cm}

\begin{abstract}
The LHCb experiment has made several measurements in $b \to s$ transitions which indicate tensions with the Standard Model predictions. Assuming the source of these tensions  to be new physics, we present new global fits to all Wilson 
coefficients which can effectively receive beyond the Standard Model contributions. While the theoretically clean ratios $R_{K^{(*)}}$ which are sensitive to lepton flavour non-universality may unambiguously establish lepton non-universal new physics in the near future, most of the other tensions with the SM in the $b \rightarrow s$ data, in particular in the angular observables of the $B\to K^* \mu\mu$ decay and in the branching ratio of the $B_s \to \phi \mu\mu$ decay, depend on the estimates of non-factorisable power corrections. Therefore, we also analyse the dependence of the new global fit on these corrections.

\end{abstract}


\end{frontmatter}

\section{Introduction}\vspace*{-0.15cm}

In the past few years, experimental measurements on several $b \to s$ transitions have indicated tensions with the Standard Model (SM) predictions at the level of $\sim2-3\sigma$ significance. 
The first and most notable tension was observed in 2013 in the angular observable $P_5^\prime$ of the $B \to K^* \mu^+ \mu^-$ decay in the $[4.30,8.68]$ GeV$^2$ bin at LHCb~\cite{Aaij:2013qta} with 1 fb$^{-1}$ of data and was confirmed by the same experiment in finer bins with 3 fb$^{-1}$ of data~\cite{Aaij:2015oid}. The branching ratio of the $B_s \to \phi \mu^+ \mu^-$ decay was also measured by  LHCb to have $\sim3\sigma$ tension with the SM prediction~\cite{Aaij:2015esa}. Other tensions with a significance of $2.2-2.6\sigma$ have been found in the ratios $R_K$ and 
$R_{K^{*}}$ by  LHCb~\cite{Aaij:2014ora,Aaij:2017vbb}. If confirmed they would establish the breaking of lepton flavour universality. 
A very promising feature of  these tensions is that they can all be explained by a common  new physics (NP) contribution~\cite{Hurth:2014vma,Altmannshofer:2017yso, Capdevila:2017bsm,DAmico:2017mtc,Hiller:2017bzc,Geng:2017svp,Ciuchini:2017mik,Hurth:2017hxg}
which enforces the idea that the present anomalies in the $b \to s$ data are due to NP.

The $P_5^\prime$ observable has also been measured by the Belle, ATLAS and CMS experiments~\cite{Abdesselam:2016llu,Aaboud:2018krd,Sirunyan:2017dhj} with larger experimental errors. The confirmation of the anomaly by the former two experiments
makes it most unlikely that the tension stems from  statistical fluctuations. It is more plausible that the tension is due to NP effects or underestimated hadronic contributions~\cite{Jager:2012uw,Hurth:2013ssa,Mahmoudi:2014mja,Lyon:2014hpa,Gratrex:2015hna}.
The significance of the tensions in the angular observables such as $P_5^\prime$ depends on the estimation of the size of the hadronic uncertainties which has not yet been completely resolved~\cite{Hurth:2013ssa,Hurth:2014zja,Hurth:2016fbr,Mahmoudi:2016mgr}.
The dominant hadronic effects in the $B\to K^* \ell^+ \ell^-$ decay is due to the matrix elements of the electromagnetic ($O_7$) and the semileptonic operators ($O_{9,10}$) described in terms of the form factors which
are under rather good control. However, another source of the hadronic contributions  is from the non-local effects arising from the matrix elements of the four-quark operators ($O_{1-6}$) and the chromomagnetic operator ($O_8$). 
The leading contributions at low $q^2$ can be calculated within the QCD factorisation (QCDf) approach~\cite{Beneke:2001at,Beneke:2004dp}, but the  subleading non-factorisable power corrections are difficult to estimate. 
Usually, the size of these corrections is assumed to be a small factor (10-20\%) of the leading QCDf amplitude.

Recently a number of methods have been suggested to estimate  the power corrections using the light-cone sum rule (LCSR) method~\cite{Khodjamirian:2010vf} or  considering the analyticity structure of the amplitudes~\cite{Bobeth:2017vxj}. An empirical model has also been proposed describing the hadronic resonances via Breit-Wigner amplitudes~\cite{Blake:2017fyh}. 
The approach of Ref.~\cite{Bobeth:2017vxj} which builds upon the work of Refs.~\cite{Khodjamirian:2010vf,Khodjamirian:2012rm}
offers the most promising outlook to give a precise estimate and hence clarify whether the source of the tensions in $B\to K^* \mu^+ \mu^-$ is due to NP contributions or hadronic effects.

There is also another approach to this issue  presented in the literature: One derives a general ansatz for the power corrections compatible with their analyticity structure. The unknown parameters of this ansatz are then 
fitted to the relevant data on $B\to K^* \mu^+ \mu^-$  
and $B \to K^* \gamma$~\cite{Ciuchini:2015qxb,Chobanova:2017ghn,Neshatpour:2017qvi}. 
With the help of  Wilks' theorem, it is then possible to make a statistical comparison of the hadronic parameters fit and the NP fit of the Wilson coefficients.

In contrast, $R_K$ and $R_{K^*}$ are theoretically very clean observables since the hadronic contributions cancel out in the ratios~\cite{Hiller:2003js,Bordone:2016gaq}. 
As a consequence, in case further experimental results confirm the observed tensions in the ratios, it can be unequivocally inferred that these anomalies are due to NP.
The different status of the  observables with respect to the hadronic corrections suggests separate studies of these two sets of observables. In Ref.~\cite{Hurth:2017hxg} it has been  shown that the NP analyses
of these two sets are less coherent than often stated, but compatible with each other at least at the 2$\sigma$ level.

It was shown in Ref.~\cite{Hurth:2017hxg} that even with a small part of the 50 fb$^{-1}$ dataset of the LHCb upgrade{, $R_K$ and $R_{K^*}$ can establish lepton flavour universality violating NP in the near future, but more such theoretically clean ratios are needed to differentiate between the various NP hypotheses. 
Moreover, in case NP is established via the ratios, it also indirectly enforces the new physics explanation of the anomalies in the $B\to K^* \mu^+ \mu^-$ and $B_s \to \phi \mu^+ \mu^-$ decays.

Keeping these issues in mind, we present our new global fits of all $b \to s$ data including the ratios $R_K$ and $R_{K^*}$ in the following.
Since we are allowing for lepton non-universal NP effects in the Wilson coefficients and also due to the theoretical cleanliness of the $R_{K^{(*)}}$ ratios compared to the other $b\to s$ data, it is expected 
that the ratios  are the dominating observables for the NP significance in the global fit.

In a model-independent analysis, there is \emph{a priori} no reason to assume that the $b \to s \ell^+ \ell^-$ anomalies are due to only one type of NP contribution. Thus, we have performed  multi-dimensional fits where all the relevant Wilson coefficients are taken into account.

\section{BR($B_s \to \mu^+ \mu^-$) and scalar operators}\label{sec:Bsmumu}\vspace*{-0.15cm}
A complete NP scenario  may include several new particles and can have extended Higgs sector, affecting the Wilson coefficients $C_{7\cdots10}$ and involving scalar and pseudoscalar contributions. Hence within a model-independent analysis it is reasonable to include all relevant Wilson coefficients in the global fit. The meaning of relevance 
in this context will be clarified below~\footnote{In an effective field theory approach, there are also other issues like scale dependence and dependence on the basis of operators. Their impact should be analysed in a case by case study.}.

The NP contributions to the scalar and pseudoscalar Wilson 
coefficients $C_{Q_{1,2}}$ (see e.g. Ref.~\cite{Mahmoudi:2012un} for the definition of the corresponding operators) are often assumed to be severely constrained from the data on BR($B_s \to \mu^+ \mu^-$).
Although this is true in many NP scenarios, in general, there is a degeneracy between $C_{10}$ and $C_{Q_2}$ allowing for large
contributions to these Wilson coefficients.

\begin{figure}[h!]
\centering
\includegraphics[width=0.40\textwidth]{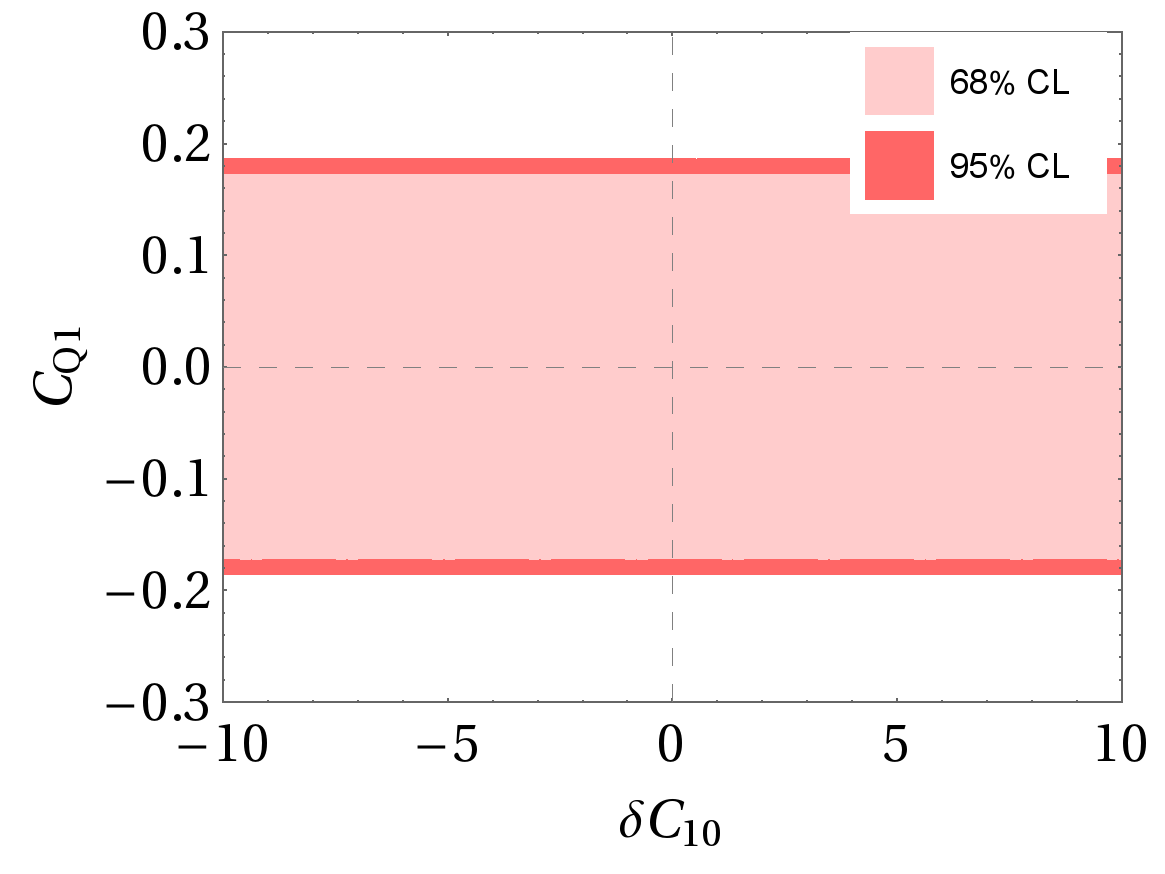}
\includegraphics[width=0.40\textwidth]{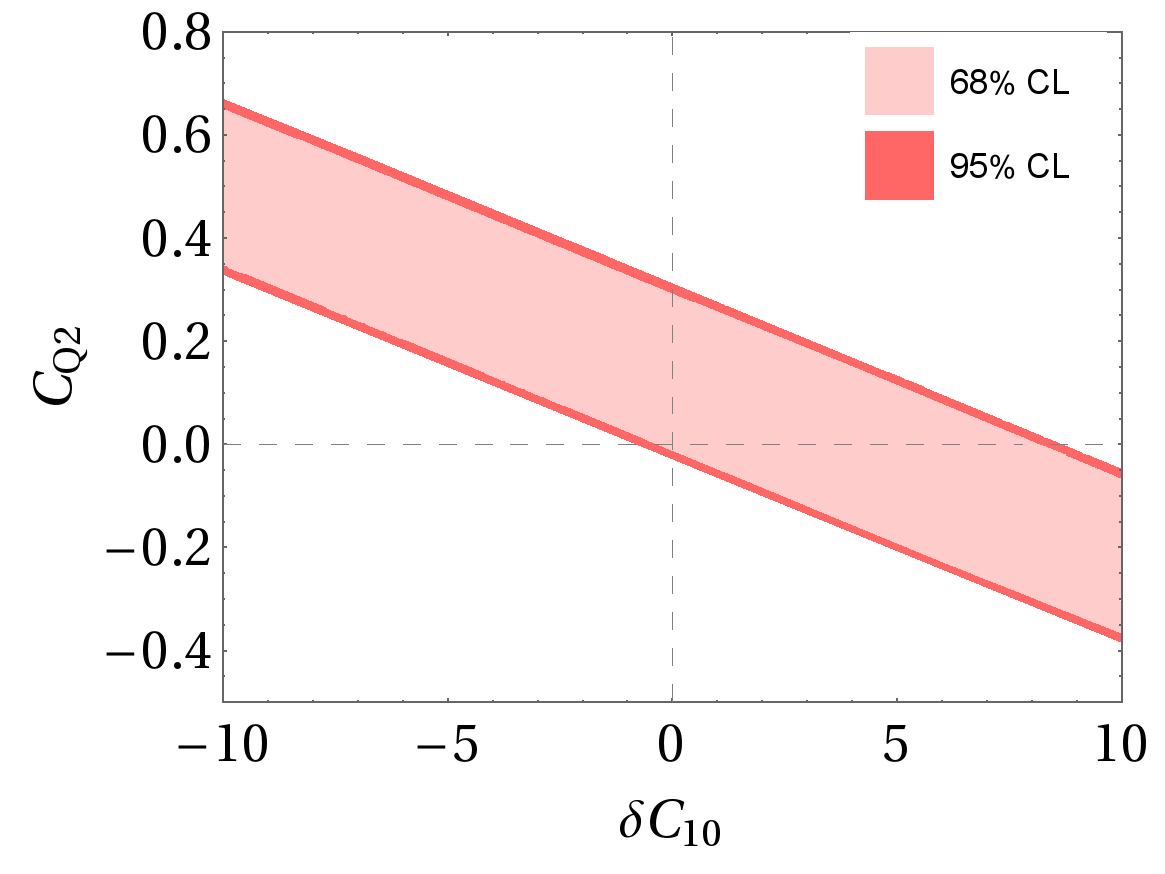}
\caption{Two-dimensional projection of the three operator fit to $C_{10}$, $C_{Q_{1}}$ and $C_{Q_{2}}$. The (light) red contours correspond to the (68) 95\% C.L. regions.
\label{Fig:C10CQ1CQ2_3op}}\vspace*{-0.4cm}
\end{figure}

\mbox{Varying}  $C_{Q_{1}}, C_{Q_{2}}$ and $C_{10}$ independently, the fit to only BR($B_s \to \mu^+ \mu^-$), leads  to the results given in Fig.~\ref{Fig:C10CQ1CQ2_3op} where two-dimensional projections  of the three operator fits are shown. Although $C_{Q_1}$ gets strongly constrained to values between $\pm0.2$, the Wilson coefficients $C_{Q_2}$ and $C_{10}$ can 
receive large contributions as they cancel out each other.

\begin{figure}[h!]
\centering
\includegraphics[width=0.40\textwidth]{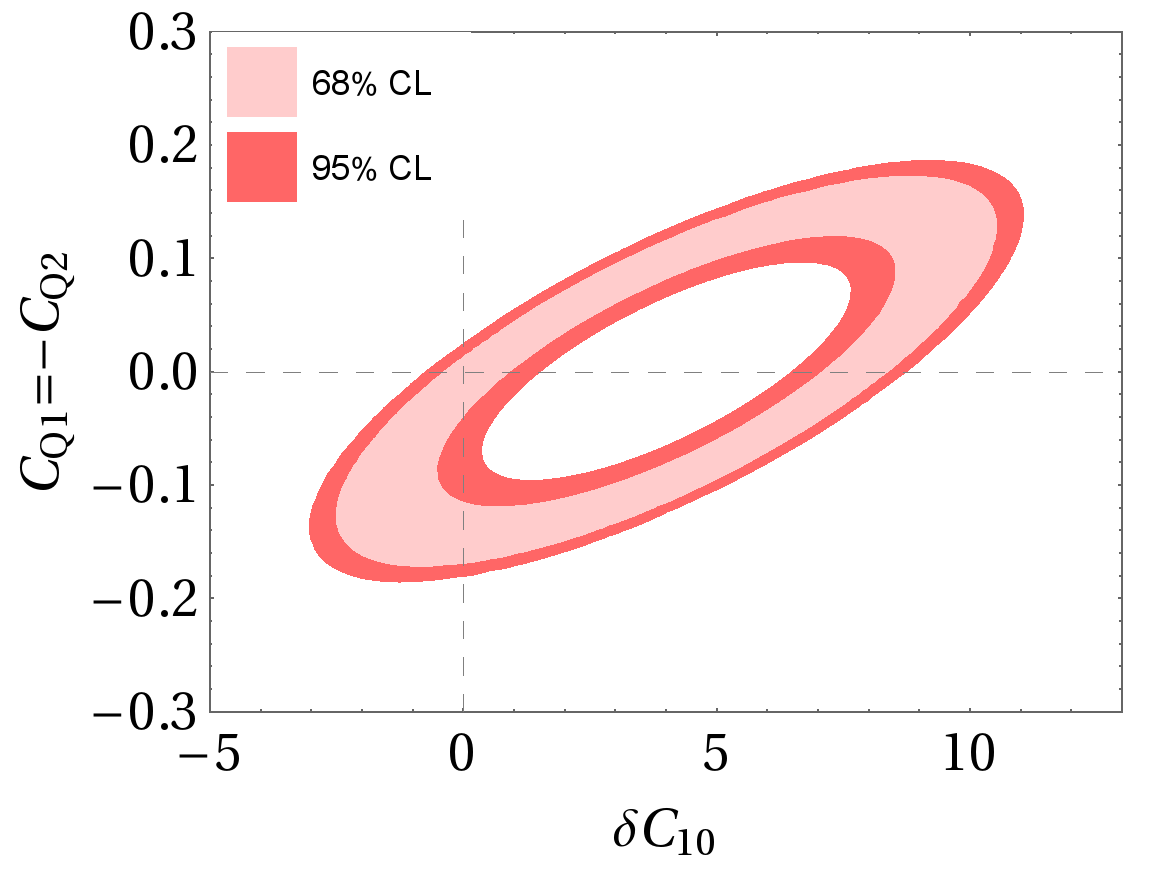}
\caption{Two operator fit to the BR($B_s \to \mu^+ \mu^-$) with NP contributions in  $C_{10}$ and $C_{Q_{1}}=-C_{Q_{2}}$.
The (light) red contours correspond to the (68) 95\% C.L. regions. 
\label{Fig:C10CQ_2op}}\vspace*{-0.3cm}
\end{figure}

We have also studied the case where  the relation $C_{Q_1}=-C_{Q_2}(\equiv C_Q)$ holds. It can be derived for example  in the SMEFT framework with a SM Higgs~\cite{Alonso:2014csa}. 
The findings can be  seen in Fig.~\ref{Fig:C10CQ_2op}. The degeneracy among $C_{10}$ and $C_{Q_2}$ is removed and we have strong constraints both for the scalar and pseudoscalar coefficients. In fact, the scalar and pseudoscalar contributions are limited between $\pm 0.2$, but $\delta C_{10}$ can still take large values.

Consequently, the branching ratio of the $B_s \to \mu^+ \mu^-$
cannot be used for the purpose of  neglecting  the scalar and pseudoscalar operators and of simultaneously putting strong constraints on $C_{10}$ in generic NP scenarios (see also Refs.~\cite{Beaujean:2015gba,Altmannshofer:2017wqy} 
for how the various $b\to s$ observables are relevant for constraining the scalar as well as tensorial Wilson coefficients).

\section{Global fits with all relevant operators}\label{sec:allWilson}\vspace*{-0.15cm}

We have expanded our study to include NP in the global fit 
to all $b \to s \ell^+ \ell^-$ data (see Ref.~\cite{Arbey:2018ics} for the considered observables) 
from all the relevant Wilson coefficients $C_7, C_8, C_9^{\ell}, C_{10}^{\ell}, C_{Q_1}^\ell, C_{Q_2}^\ell$
(assuming lepton flavours to be $\ell=e,\mu$) which, when  
considering their chirality-flipped counterparts, results in 20 Wilson coefficients.
The theoretical correlations and errors in our global fits are calculated by  {\tt SuperIso v4.0}~\cite{Mahmoudi:2007vz,Mahmoudi:2008tp}.
As a starting point, we have added  a $10\%$ error to the non-factorisable QCDf amplitude which in many cases  (depending on the observable) represents not more than a third of the complete QCDf amplitude.

In a first step, one should consider fits to one single Wilson coefficient at a time. In Ref.~\cite{Arbey:2018ics} we have found that the most favoured scenario in the one-dimensional fit is when there is NP in $C_9^{\mu}$ with a significance of $5.8\sigma$. {\it The best fit value shows about 25\% reduction in this Wilson coefficient compared to one in the SM.} In the chiral basis 
(see Ref.~\cite{Hurth:2017hxg} for the precise definition)  the scenario with NP in $C_{LL}^\mu$ (indicating left-handed lepton and quark currents) has also a large significance of $5.8\sigma$. These results are in agreement with several recent fits with similar {sets} of observables (see e.g. Refs.~\cite{Capdevila:2017bsm,DAmico:2017mtc}).

In addition, we have made a Wilks' test and considered a NP contribution only in $C_9^\mu$ and then expanded the fit, varying simultaneously $2, 6, 10$ and $20$ Wilson coefficients.
The results of the fits including $C_9^\mu$ are given in Table~\ref{tab:NPimprovements} where the last column corresponds to the improvement in comparison to the previous set of Wilson coefficients, obtained using the Wilks' theorem.  The best fit values of the one- and the multi-dimensional fits can be found in Ref.~\cite{Arbey:2018ics}. 
The SM pull decreases with the number of Wilson coefficients which is due to the increase in the number of degrees of freedom.
If there is no real improvement of the fit,  the increase of the number of degrees of freedom due to a larger number of Wilson coefficients results in a reduced pull with the SM.  This is confirmed by the Wilks' test, which explicitly shows  that adding Wilson coefficients to the ``$C_9^\mu$ only'' set does not bring a significant improvement.

\begin{table}[!t]
\begin{center}
\scalebox{0.75}{
\begin{tabular}{c|c|c|c|c}
Set of WC & param. & $\chi^2_{\rm min}$ & Pull$_{\rm SM}$ & Improvement\\
\hline
SM & 0 & 118.8 & - & -\\
$C_9^{\mu}$ & 1 & 85.1 & $5.8\sigma$ & $5.8\sigma$\\
$C_9^{(e,\mu)}$ & 2 & 83.9 & $5.6\sigma$ & $1.1\sigma$\\
$C_7,C_8,C_9^{(e,\mu)},C_{10}^{(e,\mu)}$ & 6 & 81.2 & $4.8\sigma$ & $0.5\sigma$\\
All non-primed WC & 10 (8) & 81.0 & $4.1\, (4.5)\sigma$ & $0.0\, (0.1)\sigma$\\
All WC (incl. primed) & 20 (16) & 70.2 & $3.6\, (4.1)\sigma$ & $0.9\, (1.2)\sigma$\\
\end{tabular}
}
\caption{The $\chi^2_{\rm min}$ values when varying different Wilson coefficients. 
In the last column the significance of the improvement of the fit compared to the scenario of the previous line is given.
The numbers in the parenthesis correspond to removing $C_{Q_{1,2}}^{e \, \prime}$ from the relevant fits.
\label{tab:NPimprovements}}\vspace*{-0.6cm}
\end{center} 
\end{table}

Since it is very likely that a UV-complete NP model contains several  new particles influencing various Wilson coefficients, we have varied all the 20 Wilson coefficients simultaneously. The best fit values are given in  
Table~\ref{tab:allobs_20D_C78910C12primes}. Several Wilson coefficients are only loosely constrained. The reason for this is the large number of free parameters relative to the number of observables in the fit, but also the lack of observables with significant sensitivity to these Wilson coefficients. 

Potentially large contributions to the electron Wilson coefficient are indicated by the best fit values. This is a remarkable  
phenomenon since the very few measurements of electron modes are much more SM-like than the analogous muon contributions.  The observables $R_{K^{(*)}}$ are the driving force for this. 
In particular, missing experimental results on $B_s \to e^+ e^-$ leads to large contribution in the electron scalar coefficients. As a consequence, these coefficients are undetermined in our 20-dimensional fit while the muon scalar and pseudoscalar contributions can only have very small values. 
Finally, we  have found that the best fit value  of $C_9^\mu$ is even smaller than the one in the one-dimensional fit, namely with a 35\% reduction compared to its SM value.

A comment about the number of degrees of freedom is in order here. As can be seen from Table~\ref{tab:allobs_20D_C78910C12primes}, 
$C_{Q_{1,2}}^{e \, (\prime)}$ are ``undetermined'' due to their very large uncertainties. We checked explicitly how the variation of order one in each Wilson coefficient affects the $\chi^2$, which confirmed that the four $C_{Q_{1,2}}^{e \, (\prime)}$ coefficients have a negligible impact on the fit, { i.e. for each coefficient $|\delta C_i| \sim 1$ implies $|\delta \chi^2| < 1$}. Therefore, one can define an effective number of degrees of freedom in which the insensitive coefficients are not counted~\footnote{We emphasise here that there is not a generally accepted statistical method for removing irrelevant 
parameters from the fit.}.  The results with ``relevant parameters only'' are shown in parentheses in 
Tables~\ref{tab:NPimprovements} and~\ref{tab:allobs_20D_C78910C12primes}: As a result of the full fit including all the relevant Wilson coefficients, we have obtained  a total pull of 4.1$\sigma$ with the SM hypothesis (assuming 10\% error for the non-factorisable power corrections).

\begin{table}[!t]
\begin{center}
\setlength\extrarowheight{3pt}
\scalebox{0.8}{
\begin{tabular}{|c|c|c|c|}
\hline 
  \multicolumn{4}{|c|}{All observables  with $\chi^2_{\rm SM}=118.8$} \\ 
  \multicolumn{4}{|c|}{($\chi^2_{\rm min}=70.2;\; {\rm Pull}_{\rm SM}=3.6\,(4.1)\sigma$)} \\ 
\hline \hline
\multicolumn{2}{|c}{$\delta C_7$} &  \multicolumn{2}{|c|}{$\delta C_8$}\\ 
% \hline
\multicolumn{2}{|c}{$-0.01  \pm 0.05 $} & \multicolumn{2}{|c|}{$ 0.89 \pm 0.81 $}\\ 
\hline 
%%%%%%%%%%%%%%%%%%%%%%%%%%%
\multicolumn{2}{|c}{$\delta C_7^\prime$} &  \multicolumn{2}{|c|}{$\delta C_8^\prime$}\\ 
% \hline
\multicolumn{2}{|c}{$ 0.01  \pm 0.03 $} & \multicolumn{2}{|c|}{$ -1.70 \pm 0.46 $}\\ 
\hline \hline
%%%%%%%%%%%%%%%%%%%%%%%%%%%
$\delta C_{9}^{\mu}$ & $\delta C_{9}^{e}$ & $\delta C_{10}^{\mu}$ & $\delta C_{10}^{e}$ \\
% \hline
$ -1.40 \pm 0.26 $ & $ -4.02  \pm 5.58 $ & $-0.07 \pm 0.28 $ & $ 1.32  \pm 2.02 $  \\
\hline 
%%%%%%%%%%%%%%%%%%%%%%%%%%%
$\delta C_{9}^{\prime \mu}$ & $\delta C_{9}^{\prime e}$ & $\delta C_{10}^{\prime \mu}$ & $\delta C_{10}^{\prime e}$ \\
$ 0.23 \pm 0.65 $ & $-1.10  \pm 5.98 $ & $ -0.16 \pm 0.38 $ & $ 2.70  \pm 2 $ \\
\hline \hline
%%%%%%%%%%%%%%%%%%%%%%%%%%%
$C_{Q_{1}}^{\mu}$ & $C_{Q_{1}}^{e}$ & $C_{Q_{2}}^{\mu}$ & $C_{Q_{2}}^{e}$ \\ 
% \hline
$-0.13  \pm 1.86 $ & undetermined & $-0.05  \pm 0.58 $ & undetermined \\
\hline 
%%%%%%%%%%%%%%%%%%%%%%%%%%%
$C_{Q_{1}}^{\prime \mu}$ & $C_{Q_{1}}^{\prime e}$ & $C_{Q_{2}}^{\prime \mu}$ & $C_{Q_{2}}^{\prime e}$ \\ 
$ 0.01  \pm 1.87 $ & undetermined & $-0.18  \pm 0.62 $ & undetermined \\
\hline
%%%%%%%%%%%%%%%%%%%%%%%%%%%
%%%%%%%%%%%%%%%%%%%%%%%%%%%
\end{tabular}
}
\caption{Best fit values for the 20 operator global fit to the $b \to s$ data, assuming 10\% error for the power corrections.
${\rm Pull}_{\rm SM}$ in the parenthesis corresponds to considering the effective number of degrees of freedom~(16) when giving the significance.
\label{tab:allobs_20D_C78910C12primes}}\vspace*{-0.6cm}
\end{center} 
\end{table}
\begin{table}[!h]
\begin{center}
\setlength\extrarowheight{3pt}
\scalebox{0.8}{
\begin{tabular}{|c|c|c|c|}
\hline 
  \multicolumn{4}{|c|}{All observables  with $\chi^2_{\rm SM}=118.8$} \\ 
  \multicolumn{4}{|c|}{($\chi^2_{\rm min}=74.0;\; {\rm Pull}_{\rm SM}=3.8\,(4.1)\sigma$)} \\ 
\hline \hline
\multicolumn{2}{|c}{$\delta C_7$} &  \multicolumn{2}{|c|}{$\delta C_8$}\vspace*{-0.1cm}\\ 
% \hline
\multicolumn{2}{|c}{$ 0.00  \pm 0.05 $} & \multicolumn{2}{|c|}{$ 0.81 \pm 0.80 $}\\ 
\hline 
%%%%%%%%%%%%%%%%%%%%%%%%%%%
\multicolumn{2}{|c}{$\delta C_7^\prime$} &  \multicolumn{2}{|c|}{$\delta C_8^\prime$}\vspace*{-0.1cm}\\ 
% \hline
\multicolumn{2}{|c}{$ 0.01  \pm 0.03 $} & \multicolumn{2}{|c|}{$ -1.74 \pm 0.46 $}\\ 
\hline \hline
%%%%%%%%%%%%%%%%%%%%%%%%%%%
$\delta C_{9}^{\mu}$ & $\delta C_{9}^{e}$ & $\delta C_{10}^{\mu}$ & $\delta C_{10}^{e}$ \vspace*{-0.1cm}\\
% \hline
$ -1.44 \pm 0.26 $ & $ -4.99  \pm 5.92 $ & $-0.01 \pm 0.29 $ & $ -0.86  \pm 3.77 $  \\
\hline 
%%%%%%%%%%%%%%%%%%%%%%%%%%%
$\delta C_{9}^{\prime \mu}$ & $\delta C_{9}^{\prime e}$ & $\delta C_{10}^{\prime \mu}$ & $\delta C_{10}^{\prime e}$ \vspace*{-0.1cm}\\
$ 0.18 \pm 0.65 $ & $-2.02  \pm 9.43 $ & $ -0.15 \pm 0.38 $ & $ -0.33  \pm 2.09 $ \\
\hline \hline
%%%%%%%%%%%%%%%%%%%%%%%%%%%
\multicolumn{2}{|c}{$C_Q^{\mu}\; (C_{Q_1}^{\mu}=-C_{Q_2}^{\mu})$} &  \multicolumn{2}{|c|}{$ C_Q^{e}\; (C_{Q_1}^{e}=-C_{Q_2}^{e})$}\vspace*{-0.1cm}\\ 
% \hline
% \multicolumn{2}{|c}{$-0.01   \pm 0.18 $} & \multicolumn{2}{|c|}{$-0.08 \pm 4.91 $}\\ 
\multicolumn{2}{|c}{$-0.01   \pm 0.18 $} & \multicolumn{2}{|c|}{undetermined}\\ 
\hline 
%%%%%%%%%%%%%%%%%%%%%%%%%%%
\multicolumn{2}{|c}{$C_Q^{\prime \mu}\; (C_{Q_1}^{\prime \mu}=+C_{Q_2}^{\prime \mu})$} &  \multicolumn{2}{|c|}{$ C_Q^{\prime e}\; (C_{Q_1}^{\prime e}=+C_{Q_2}^{\prime e})$}\vspace*{-0.1cm}\\ 
% \hline
% \multicolumn{2}{|c}{$-0.15  \pm 0.13 $}  & \multicolumn{2}{|c|}{$ 0.13 \pm 3.03 $}\\ 
\multicolumn{2}{|c}{$-0.15  \pm 0.13 $}  & \multicolumn{2}{|c|}{undetermined}\\ 
\hline
%%%%%%%%%%%%%%%%%%%%%%%%%%%
%%%%%%%%%%%%%%%%%%%%%%%%%%%
\end{tabular}
}
\caption{Best fit values for the 16 operator global fit to the $b \to s$ data, considering the SMEFT relations between $C_{Q_1}$ and  $C_{Q_2}$
and assuming 10\% error for the power corrections.
${\rm Pull}_{\rm SM}$ in the parenthesis corresponds to the effective number of degrees of freedom~(14). 
\label{tab:allobs_16D_SMEFT}} \vspace*{-0.4cm}
\end{center} 
\end{table}

\begin{table}[!t]
\begin{center}
\setlength\extrarowheight{3pt}
\scalebox{0.8}{
\begin{tabular}{|c|c|c|c|}
\hline 
  \multicolumn{4}{|c|}{All observables} \\ 
%   \multicolumn{4}{|c|}{10\% pc:\;\; $\chi^2_{\rm SM}=118.8 ,\;\; \chi^2_{\rm min}=70.2;\; {\rm Pull}_{\rm SM}=3.6\sigma\,(4.1\sigma)$} \\ \cline{2-3}
  \multicolumn{4}{|c|}{30\% pc:\;\; $\chi^2_{\rm SM}=115.5 ,\;\; \chi^2_{\rm min}=65.8;\; {\rm Pull}_{\rm SM}=3.7\sigma\,(4.2\sigma)$} \\ \cline{2-3}
%   \multicolumn{4}{|c|}{40\% pc:\;\; $\chi^2_{\rm SM}=114.2 ,\;\; \chi^2_{\rm min}=64.8;\; {\rm Pull}_{\rm SM}=3.6\sigma\,(4.2\sigma)$} \\ \cline{2-3}
%   \multicolumn{4}{|c|}{50\% pc:\;\; $\chi^2_{\rm SM}=113.0 ,\;\; \chi^2_{\rm min}=64.1;\; {\rm Pull}_{\rm SM}=3.6\sigma\,(4.1\sigma)$} \\ \cline{2-3}
  \multicolumn{4}{|c|}{60\% pc:\;\; $\chi^2_{\rm SM}=111.8 ,\;\; \chi^2_{\rm min}=63.6;\; {\rm Pull}_{\rm SM}=3.5\sigma\,(4.1\sigma)$} \\ \cline{2-3}
  \multicolumn{4}{|c|}{100\% pc:\;\; $\chi^2_{\rm SM}=107.3 ,\;\; \chi^2_{\rm min}=62.0;\; {\rm Pull}_{\rm SM}=3.3\sigma\,(3.8\sigma)$} \\ 
\hline \hline
\multicolumn{2}{|c}{$\delta C_7$} &  \multicolumn{2}{|c|}{$\delta C_8$} \vspace*{-0.1cm}\\ 
% \multicolumn{2}{|c}{$ -0.01	\pm	0.05$} & \multicolumn{2}{|c|}{$ 0.89	\pm	0.81$} \vspace*{-0.2cm}\\           
\multicolumn{2}{|c}{$ -0.01	\pm	0.05$} & \multicolumn{2}{|c|}{$ 0.89	\pm	0.84$} \vspace*{-0.2cm}\\ 
% \multicolumn{2}{|c}{$ -0.01	\pm	0.05$} & \multicolumn{2}{|c|}{$ 0.89	\pm	0.85$} \vspace*{-0.2cm}\\ 
% \multicolumn{2}{|c}{$ -0.01	\pm	0.05$} & \multicolumn{2}{|c|}{$ 0.90	\pm	0.79$} \vspace*{-0.2cm}\\ 
\multicolumn{2}{|c}{$ -0.02	\pm	0.05$} & \multicolumn{2}{|c|}{$ 0.89	\pm	0.79$} \vspace*{-0.2cm}\\ 
\multicolumn{2}{|c}{$ -0.02	\pm	0.05$} & \multicolumn{2}{|c|}{$ 0.90	\pm	0.85$} \\ 
\hline 
%%%%%%%%%%%%%%%%%%%%%%%%%%%
\multicolumn{2}{|c}{$\delta C_7^\prime$} &  \multicolumn{2}{|c|}{$\delta C_8^\prime$} \vspace*{-0.1cm}\\ 
% \multicolumn{2}{|c}{$ 0.01	\pm	0.03$} & \multicolumn{2}{|c|}{$ -1.70	\pm	0.46$} \vspace*{-0.2cm}\\ 
\multicolumn{2}{|c}{$ 0.01	\pm	0.03$} & \multicolumn{2}{|c|}{$ -1.68	\pm	0.49$} \vspace*{-0.2cm}\\ 
% \multicolumn{2}{|c}{$ 0.02	\pm	0.03$} & \multicolumn{2}{|c|}{$ -1.67	\pm	0.51$} \vspace*{-0.2cm}\\ 
% \multicolumn{2}{|c}{$ 0.01	\pm	0.04$} & \multicolumn{2}{|c|}{$ -1.65	\pm	0.51$} \vspace*{-0.2cm}\\ 
\multicolumn{2}{|c}{$ 0.01	\pm	0.04$} & \multicolumn{2}{|c|}{$ -1.63	\pm	0.52$} \vspace*{-0.2cm}\\ 
\multicolumn{2}{|c}{$ 0.01	\pm	0.05$} & \multicolumn{2}{|c|}{$ -1.53	\pm	0.61$} \\ 
\hline \hline
%%%%%%%%%%%%%%%%%%%%%%%%%%%
$\delta C_{9}^{\mu}$ & $\delta C_{9}^{e}$ & $\delta C_{10}^{\mu}$ & $\delta C_{10}^{e}$ \vspace*{-0.1cm}\\
% $ -1.40	\pm	0.26	 $ & $ -4.02	\pm	5.58	 $ & $ -0.07	\pm	0.28	 $ & $ 1.32	\pm	2.02	 $  \vspace*{-0.2cm} \\
$ -1.43	\pm	0.27	 $ & $ -4.07	\pm	5.38	 $ & $ -0.06	\pm	0.30	 $ & $ 1.34	\pm	1.98	 $  \vspace*{-0.2cm} \\
% $ -1.43	\pm	0.28	 $ & $ -4.08	\pm	5.37	 $ & $ -0.05	\pm	0.31	 $ & $ 1.34	\pm	1.99	 $  \vspace*{-0.2cm} \\
% $ -1.43	\pm	0.28	 $ & $ -4.08	\pm	5.22	 $ & $ -0.04	\pm	0.32	 $ & $ 1.34	\pm	1.77	 $  \vspace*{-0.2cm} \\
$ -1.43	\pm	0.28	 $ & $ -4.14	\pm	3.02	 $ & $ -0.04	\pm	0.33	 $ & $ 1.34	\pm	1.76	 $  \vspace*{-0.2cm} \\
$ -1.40	\pm	0.29	 $ & $ -3.95	\pm	4.53	 $ & $ -0.01	\pm	0.37	 $ & $ 1.37	\pm	1.80	 $  \\
\hline 
%%%%%%%%%%%%%%%%%%%%%%%%%%%
$\delta C_{9}^{\prime \mu}$ & $\delta C_{9}^{\prime e}$ & $\delta C_{10}^{\prime \mu}$ & $\delta C_{10}^{\prime e}$ \vspace*{-0.1cm}\\
% $ 0.23	\pm	0.65	 $ & $ -1.10	\pm	5.98	 $ & $ -0.16	\pm	0.38	 $ & $ 2.69	\pm	2.01	 $  \vspace*{-0.2cm} \\
$ 0.22	\pm	0.71	 $ & $ -1.10	\pm	5.78	 $ & $ -0.16	\pm	0.40	 $ & $ 2.68	\pm	1.98	 $  \vspace*{-0.2cm} \\
% $ 0.22	\pm	0.75	 $ & $ -1.09	\pm	5.80	 $ & $ -0.15	\pm	0.41	 $ & $ 2.67	\pm	1.98	 $  \vspace*{-0.2cm} \\
% $ 0.21	\pm	0.77	 $ & $ -1.10	\pm	5.58	 $ & $ -0.15	\pm	0.40	 $ & $ 2.67	\pm	1.76	 $  \vspace*{-0.2cm} \\
$ 0.21	\pm	0.79	 $ & $ -1.02	\pm	3.40	 $ & $ -0.14	\pm	0.41	 $ & $ 2.67	\pm	1.75	 $  \vspace*{-0.2cm} \\
$ 0.28	\pm	0.88	 $ & $ -1.24	\pm	5.05	 $ & $ -0.11	\pm	0.45	 $ & $ 2.64	\pm	1.84	 $  \\
\hline \hline
%%%%%%%%%%%%%%%%%%%%%%%%%%%
$C_{Q_{1}}^{\mu}$ & $C_{Q_{1}}^{e}$ & $C_{Q_{2}}^{\mu}$ & $C_{Q_{2}}^{e}$ \vspace*{-0.1cm}\\ 
% $ -0.13	\pm	1.86	 $ & $ -4.87	\pm	6.76	 $ & $ -0.05	\pm	0.58	 $ & $ 0.66	\pm	21.01	 $  \vspace*{-0.2cm} \\
% $ -0.11	\pm	2.03	 $ & $ -4.72	\pm	7.86	 $ & $ -0.07	\pm	0.36	 $ & $ 1.27	\pm	18.91	 $  \vspace*{-0.2cm} \\
% $ -0.11	\pm	1.46	 $ & $ -4.28	\pm	14.88	 $ & $ -0.07	\pm	0.37	 $ & $ 2.32	\pm	25.62	 $  \vspace*{-0.2cm} \\
% $ -0.12	\pm	0.15	 $ & $ -3.97	\pm	6.94	 $ & $ -0.07	\pm	0.20	 $ & $ 2.79	\pm	8.78	 $  \vspace*{-0.2cm} \\
% $ -0.15	\pm	1.00	 $ & $ -3.69	\pm	7.22	 $ & $ -0.05	\pm	0.36	 $ & $ 3.13	\pm	8.18	 $  \vspace*{-0.2cm} \\
% $ -0.08	\pm	1.04	 $ & $ -1.96	\pm	11.64	 $ & $ -0.08	\pm	0.25	 $ & $ 4.41	\pm	6.09	 $  \\
%%%
% $ -0.13	\pm	1.86	 $ &                               & $ -0.05	\pm	0.58	 $ &                                \vspace*{-0.2cm} \\
$ -0.11	\pm	2.03	 $ &                               & $ -0.07	\pm	0.36	 $ &                                \vspace*{-0.2cm} \\
% $ -0.11	\pm	1.46	 $ &          undetermined         & $ -0.07	\pm	0.37	 $ &          undetermined          \vspace*{-0.2cm} \\
% $ -0.12	\pm	0.15	 $ &                               & $ -0.07	\pm	0.20	 $ &                                \vspace*{-0.2cm} \\
$ -0.15	\pm	1.00	 $ &          undetermined         & $ -0.05	\pm	0.36	 $ &          undetermined          \vspace*{-0.2cm} \\
$ -0.08	\pm	1.04	 $ &                               & $ -0.08	\pm	0.25	 $ &                                \\
\hline 
%%%%%%%%%%%%%%%%%%%%%%%%%%%
$C_{Q_{1}}^{\prime \mu}$ & $C_{Q_{1}}^{\prime e}$ & $C_{Q_{2}}^{\prime \mu}$ & $C_{Q_{2}}^{\prime e}$ \vspace*{-0.1cm}\\ 
% $ 0.01	\pm	1.87	 $ & $ -4.87	\pm	7.01	 $ & $ -0.18	\pm	0.62	 $ & $ 0.67	\pm	20.99	 $  \vspace*{-0.2cm} \\
% $ 0.03	\pm	2.07	 $ & $ -4.72	\pm	7.89	 $ & $ -0.18	\pm	0.53	 $ & $ 1.28	\pm	18.88	 $  \vspace*{-0.2cm} \\
% $ 0.04	\pm	1.46	 $ & $ -4.28	\pm	14.90	 $ & $ -0.18	\pm	0.37	 $ & $ 2.32	\pm	25.84	 $  \vspace*{-0.2cm} \\
% $ 0.03	\pm	0.15	 $ & $ -3.97	\pm	6.95	 $ & $ -0.18	\pm	0.22	 $ & $ 2.80	\pm	8.72	 $  \vspace*{-0.2cm} \\
% $ 0.00	\pm	1.03	 $ & $ -3.69	\pm	7.28	 $ & $ -0.17	\pm	0.51	 $ & $ 3.14	\pm	8.15	 $  \vspace*{-0.2cm} \\
% $ 0.06	\pm	1.05	 $ & $ -1.97	\pm	11.62	 $ & $ -0.18	\pm	0.22	 $ & $ 4.42	\pm	5.94	 $  \\
%%%
% $ 0.01	\pm	1.87	 $ &                               & $ -0.18	\pm	0.62	 $ &                                \vspace*{-0.2cm} \\
$ 0.03	\pm	2.07	 $ &                               & $ -0.18	\pm	0.53	 $ &                                \vspace*{-0.2cm} \\
% $ 0.04	\pm	1.46	 $ &          undetermined         & $ -0.18	\pm	0.37	 $ &          undetermined          \vspace*{-0.2cm} \\
% $ 0.03	\pm	0.15	 $ &                               & $ -0.18	\pm	0.22	 $ &                                \vspace*{-0.2cm} \\
$ 0.00	\pm	1.03	 $ &          undetermined         & $ -0.17	\pm	0.51	 $ &          undetermined          \vspace*{-0.2cm} \\
$ 0.06	\pm	1.05	 $ &                               & $ -0.18	\pm	0.22	 $ &                                \\
\hline
%%%%%%%%%%%%%%%%%%%%%%%%%%%
%%%%%%%%%%%%%%%%%%%%%%%%%%%
\end{tabular}
}
\caption{ Best fit values for the 20 operator global fit to the $b \to s$ data, assuming 30, 60 and 100\% errors for the power corrections in the first, second and third row for each Wilson coefficient, respectively.
${\rm Pull}_{\rm SM}$ in the parenthesis corresponds to the effective number of degrees of freedom~(16). 
\label{tab:allobs_20D_10_100_percent}} \vspace*{-0.6cm}
\end{center} 
\end{table}

As mentioned above, the reason why the four  parameters $C_{Q_{1,2}}^{e \, (\prime)}$ are undetermined is mainly due to the lack of observables in our present global fits which could be sensitive to these parameters. It is not ruled out that in a new global fit with new observables those parameters become relevant again. There is another question within a model-independent analysis: What happens to the parameters which are zero in the yet unknown UV-theory.  In this case, we should observe that within the model-independent analysis these parameters are shown to be zero. Also, a Wilks' test should demonstrate that these parameters do not improve the fit at all. In such a situation it also makes sense to remove these parameters from the fit. 

We have also made a model-independent fit with all Wilson coefficients under the SMEFT-relations
$C_Q^{e/\mu} \equiv C_{Q_1}^{e/\mu}= - C_{Q_2}^{e/\mu}$ and $C_Q^{e/\mu,'} \equiv C_{Q_1}^{e/\mu,'}= C_{Q_2}^{e/\mu,'}$. 
The results are presented in Table~\ref{tab:allobs_16D_SMEFT}. Again the electronic scalar operators are undetermined. The SM pull of the global fit with all relevant parameters is just the same as in the previous case.

Finally, we have analysed the dependence of the global fit on the precise estimate of the non-factorisable power correction. 
As discussed in Ref.~\cite{Hurth:2016fbr},  a 20\% estimation of the non-factorisable power correction  leads
to an error of maximum 1.5\%, 3\%, 6\% respectively on the observable level in the three angular observables  $S_3$, $S_4$ and $S_5$ and the  60\%  error estimation in this framework 
leads to 17\%-20\% error in these three observables. Thus,   
changing the power correction error between 5\% and 20\% we have found no real changes in the best fit values of the various parameters and in the SM pull as expected. In Table~\ref{tab:allobs_20D_10_100_percent}, we have varied the power correction error between 30, 60  and 100\%. Surprisingly,
only the 100\% choice leads to a significant change in the SM pull. This reflects the fact that the theoretically clean ratios dominate the NP significance in the global fit.

\vspace*{-0.1cm}
\section*{Acknowledgements}\vspace*{-0.2cm}
We thank the organisers of the Capri workshop for the very fruitful  workshop and Christoph Bobeth, Danny van Dyk, Alexander Khodjamirian, Luca Silvestrini for useful discussions. TH thanks the CERN theory group for its hospitality during his regular visits to CERN where part of this work was written. 

\vspace*{-0.2cm}
\nocite{*}
\bibliographystyle{elsarticle-num}
\bibliography{CapriRefs2018}

\end{document}